\documentclass{casc}
\newcommand{\No}{{\the\textfont2
 N}\lower-0.4ex\hbox{\small \underline{$\circ$}}}
\begin{document}
{\Large
\title{Construction of Single-valued
 Solutions for Nonintegrable Systems with the Help of the Painlev\'{e} Test}
\titlerunning{Construction of Single-Valued
 Solutions for Nonintegrable Systems}
\author{Sergey Yu.~Vernov}
\authorrunning{Sergey Yu.~Vernov}
\institute{Skobeltsyn
Institute of Nuclear Physics \ of \
Moscow State University,\\  Vorob'evy
Gory,\\ 119992 Moscow, Russia\\
\email{svernov@theory.sinp.msu.ru}}
}
\maketitle                 

\begin{abstract}
\normalsize 
The Painlev\'e test is very useful to construct not
only the Laurent-series solutions but also the elliptic and trigonometric
ones. Such single-valued functions are solutions of some polynomial first
order differential equations. To find the elliptic solutions we transform
an initial nonlinear differential equation in a nonlinear algebraic system
in parameters of the Laurent-series  solutions of the initial equation.
 The number of unknowns in the obtained nonlinear system does
not depend on number of arbitrary coefficients of the used first order
equation.  In this paper we describe the corresponding algorithm, which
has been realized in REDUCE and Maple.

\end{abstract}

\section{INTRODUCTION}

The investigations of the exact special solutions of
nonintegrable systems play an important role in the study of nonlinear
physical phenomena.  There are a few methods to construct such
solutions~\cite{Weiss1,Cantos,CoMu92,Timosh,Fan,VeTish04} in terms of
 rational, hyperbolic, trigonometric or elliptic functions.
These methods use the results of the Painlev\'e test, describing behavior
of solutions in the neighbourhood of their singular points, but do not use
local solutions obtained as the Laurent series.  In 2003 R.~Conte and
M.~Musette~\cite{CoMu03} have proposed the method, which uses such
solutions.  This method constructs global single-valued solutions in two
steps. The first step is construction of the local special solutions as
the Laurent series. The second step is construction of the first order
polynomial autonomous differential equations which have the same Laurent
series solutions. The general solutions of these equations are special
solutions of the initial system. In this paper we present the computer
algebra program, which realizes this method.

\section{THE  PAINLEV\'E ANALYSIS}

When we study some mechanical problem the time is assumed to be real,
whereas the integrability of motion  equations is connected with the
behavior of their solutions as functions of complex time. Solutions of a
system of ordinary differential equations (ODE's) are regarded as analytic
functions, maybe with isolated singular points.  A singular point of a
solution is said {\it critical } (as opposed to {\it noncritical}) if the
solution is multivalued (single-valued) in its neighbourhood  and {\it
movable} if its location depends on initial conditions~\cite{Golubev1}.
 {\it The  general solution} of an ODE of order $N$ is the set of all
solutions mentioned in the existence theorem of Cauchy, i.e. determined by
the initial values.  It depends on $N$ arbitrary independent constants.
{\it A   special solution} is any solution obtained from the general
solution by giving values to the arbitrary constants.  {\it A singular
solution} is any solution which is not special, i.e. which does not belong
to the general solution.  A system of  ODE's has \textbf{\textit{ the
Painlev\'e property }} if its general solution has no movable critical
singular point~\cite{Painleve1}.

 \textbf{\textit{The Painlev\'e test}} is any
algorithm, which checks some necessary
conditions for a differential equation to have the Painlev\'e
property.
The original algorithm, developed by P.~Painlev\'e and
used by him to find all the second order ODE's with
Painlev\'e property, is known as the $\alpha$-method.
The method of S.V.~Kovalevskaya~\cite{Kova} is not as general as the
$\alpha$--method, but much more simple. The remarkable property of this
test is that it can be checked in a finite number of steps. This test can
only detect the occurrence of logarithmic and algebraic branch points.  To
date there is no general finite algorithmic method to detect the
occurrence of essential singularities\footnote{ Different variants of the
Painlev\'e test are compared in~\cite[R.~Conte paper]{Conte0}}.
  In 1980, developing the Kovalevskaya method further,
 M.J.~Ablowitz, A.~Ramani and
 H.~Segur~\cite{ARS} constructed  a new algorithm of the Painlev\'e test for
ODE's.  This algorithm appears very useful to find  solutions as a formal
Laurent series. First of all, it allows to determine the dominant behavior
of a solution in the neighbourhood  of the singular point $t_0$.  If the
solution tends to infinity as $(t-t_0)^\beta$, where $\beta$ is a negative
integer number, then substituting the Laurent series expansions one can
transform nonlinear differential equations into a system of linear
algebraic equations on coefficients of the Laurent series.
All solutions of an autonomous system depend on the parameter
$t_0$, which characterizes the singular point location. If a single-valued
solution depends on other parameters, then some coefficients of its
Laurent series have to be arbitrary and the corresponding systems have to
have zero determinants.  The numbers of such systems (named {\it
resonances} or {\it Kovalevskaya exponents}) can be determined due to the
Painlev\'e test\footnote{ In such a way we obtain solutions only as formal
series, but for some nonintegrable systems, for example, the generalized
H\'enon--Heiles system~\cite{Melkonian}, the convergence of the Laurent-
and psi-series solutions has been proved.}.

In~\cite{CoMu03} the following classical results have been used to
construct the suitable form of the first order autonomous equation:

1. The Painlev\'e theorem~\cite{Painleve1}.
   Solutions of the equation
$$
  P(y(t),y_t(t),t)=0,
$$
where $P$ is a polynomial in both $y(t)$ and  $y_t(t)\equiv
\frac{dy(t)}{dt}$ has no movable essential singular point.

2. The Fuchs theorem~\cite{Fuchs}.
If the equation
     $$
       \sum_{k=0}^{m} P_k(y(t),t)y_t^k=0,
     $$
where $P_k(y(t),t)$ are polynomials in $y(t)$ and  analytic functions in
$t$, has no critical movable singular points, then the power of $P_k(y)$
is no more than $2m-2k$, in particular, $P_m(y)$ is a constant.

Therefore, the necessary form of a polynomial autonomous first order ODE
with the single-valued general solution is
$$
\sum_{k=0}^{m}
\sum_{j=0}^{2m-2k}a_{jk}^{\vphantom{27}}\: y^j y_{t}^k=0,
\qquad a_{0m}^{\vphantom{27}}=1, \eqno(1)
$$
in which $m$ is a positive integer number and $a_{jk}$ are constants.

3. The Briot and Bouquet theorem~\cite{BriBo}.
 If the general solution of a
polynomial autonomous first order ODE is single-valued, then this
solution is either an elliptic function,  or a rational function of
$e^{\gamma x}$, $\gamma$ being some constant, or a rational function of
$x$. Note that the third case is a degeneracy of the second one, which in
its turn is a degeneracy of the first one.

\section{THE ALGORITHM AND ITS REALIZATION}

\subsection{The Laurent-Series Solutions}

To analyze the method of the Laurent series solutions construction let us
consider the generalized H\'enon--Heiles system with an additional
non-polynomial term, which is described by the Hamiltonian:
$$
H=\frac{1}{2}\Big(x_t^2+y_t^2+\lambda_1
x^2+\lambda_2 y^2\Big)+x^2y-\frac{C}{3}\:y^3+\frac{\mu}{2x^2}
$$
and the corresponding system of the motion equations:
$$
  \left\{ \begin{array}{lcl}
  \displaystyle x_{tt}^{\vphantom{7}} {}={}-\lambda_1 x
-2xy+\frac{\mu}{x^3},\\[2mm]
 \displaystyle  y_{tt}^{\vphantom{7}} {}={}-\lambda_2 y -x^2+Cy^2,
\end{array}
\right.
\eqno(2)
$$
where $x_{tt}^{\vphantom{7}}\equiv\frac{d^2x}{dt^2}$ and
$y_{tt}^{\vphantom{7}}\equiv\frac{d^2y}{dt^2}$,  $\lambda_1$, $\lambda_2$,
$\mu$ and $C$ are arbitrary numerical parameters. Note that if
$\lambda_2\neq 0$, then one can put  $\lambda_2=sign(\lambda_2)$ without
loss of generality. If $C=1$, $\lambda_1=1$, $\lambda_2=1$ and $\mu=0$,
then $(2)$ is the initial H\'enon--Heiles system~\cite{HeHe}.

The function $y$,
solution of system~$(2)$, satisfies the following fourth-order equation,
which does not include $\mu$:
$$
y_{tttt}^{\vphantom{7}}=(2C-8)y_{tt}^{\vphantom{7}}y
- (4\lambda_1+\lambda_2)y_{tt}^{\vphantom{7}}+2(C+1)y_{t}^2+
\frac{20C}{3}y^3+ (4C\lambda_1-6\lambda_2)y^2-4\lambda_1\lambda_2 y-4H.
\eqno(3)
$$
We note that the energy of the system $H$ is not an arbitrary parameter,
but a function of initial data: $y_0^{\vphantom{7}}$,
$y_{0t}^{\vphantom{7}}$,
 $y_{0tt}^{\vphantom{7}}$ and $y_{0ttt}^{\vphantom{7}}$. The
form of this function depends on $\mu$:
$$
H=\frac{1}{2}(y_{0t}^2+y_0^2)-\frac{C}{3}y_0^3+
\left(\frac{\lambda_1}{2}+y_0^{\vphantom{7}}\right)(Cy_0^2-
\lambda_2y_0^{\vphantom{7}}-y_{0tt}^{\vphantom{7}})+
\frac{(\lambda_2y_{0t}^{\vphantom{7}}+
2Cy_0^{\vphantom{7}}y_{0t}^{\vphantom{7}}-y_{0ttt}^{\vphantom{7}})^2+\mu}
{2(Cy_0^2-\lambda_2y_0^{\vphantom{7}}-y_{0tt}^{\vphantom{7}})}.
$$

This formula is correct only if $x_0^{\vphantom{7}}=Cy_0^2
-\lambda_2y_0^{\vphantom{7}}-y_{0tt}^{\vphantom{7}}\neq0$.
If $x_0^{\vphantom{7}}=0$, what is possible only at $\mu=0$, then we can
not express $x_{0t}^{\vphantom{7}}$ through $y_0^{\vphantom{7}}$,
 $y_{0t}^{\vphantom{7}}$, $y_{0tt}^{\vphantom{7}}$ and
$y_{0ttt}^{\vphantom{7}}$, so  $H$ is not a function of the initial data.
If $y_{0ttt}^{\vphantom{7}}
=2Cy_0^{\vphantom{7}}y_{0t}^{\vphantom{7}}-\lambda_2y_{0t}^{\vphantom{7}}$,
then eq. $(3)$ with an arbitrary $H$ corresponds to system $(2)$ with
$\mu=0$, in opposite case eq. $(3)$ does not correspond to system $(2)$.

The Painlev\'e test of eq.~$(3)$ gives the following
dominant behaviors and resonance structures near the singular point $t_0$:

{\bf 1.} The function $y$ tends to infinity as $b_{-2}(t-t_0)^{-2}$, where
$b_{-2}=-3$ or $b_{-2}=\frac{6}{C}$.

{\bf 2.} For $b_{-2}=-3$ ({\it Case 1}) the values of resonances are
$$r=-1,\; 10,\; (5\pm\sqrt{1-24(1+C)})/2.$$

In  {\it Case 2} ($b_{-2}=\frac{6}{C}$)
$$r=-1,\; 5,\; 5\pm\sqrt{1-48/C}.$$

The resonance $r=-1$ corresponds to an arbitrary parameter $t_0$. Other
values of $r$ determine powers of $t$ (their values are $r-2$), at which
new arbitrary parameters can appear as solutions of the linear systems
with zero determinant.  For integrability of system $(2)$ all values of
$r$ have to be integer and all systems with zero determinants have to have
solutions at any values of free parameters included in them. It is
possible only in  integrable cases.

For the search for special solutions, it is interesting to
consider such values of $C$, for which $r$ are integer
numbers either only in {\it Case 1} or only in {\it Case 2}.
If there exist  negative integer resonances, different from $r=-1$,
then such Laurent series expansion corresponds rather to singular
than general solution.
We demand that all values of $r$, but one, are nonnegative integer numbers
and all these values are different. From these conditions we obtain the
 following values of $C$: \ $C=-1$ and $C=-4/3$ ({\it Case 1}), or
$C=-16/5$, $C=-6$ and $C=-16$ ({\it Case 2},
$\alpha=\frac{1-\sqrt{1-48/C}^{\vphantom{7^4}}}{2}$), and also
$C=-2$, in which these two {\it Cases} coincide.
It has been shown
in~\cite{VernovTMF} (for $\mu=0$) and~\cite{VeTish04} (for
 an arbitrary value of  $\mu$) that
single-valued three-parameter special solutions exist in two
nonintegrable cases: $C = -16/5$ and $C = -4/3$ ($\lambda_1$ and 
$\lambda_2$ are arbitrary).

When the resonance structure is known it is easy to write the computer
algebra program, which finds the Laurent series solutions with an
arbitrary accuracy.  For example, we have found 65 coefficients of the
Laurent series for both above-mentioned values of $C$, the sizes of the
corresponding output files are about 10 Mb.

\subsection{Two Methods for Construction of Global Single-Valued Solutions}

We have found local single-valued solutions. Of course, the existence of
local single-valued solutions is a necessary, but not a sufficient
condition for the existence of global ones, because solutions, which are
single-valued in the  neighbourhood  of one singular point, can  be
multivalued in the  neighbourhood  of another singular point. So, we can
only assume that global three-parameter solutions are single-valued. If we
assume this and moreover that these solutions are elliptic functions (or
some degenerations of them), then we can seek them as solutions of some
polynomial first order equations.

The classical method to find special analytic solutions for the
generalized H\'enon--Heiles system is the following:

1) Transform system (2) into  eq.~$(3)$.

2) Assume that $y$ satisfies some  first order equation, substitute this
equation in (3) and obtain a nonlinear algebraic system.

3) Solve the obtained system.

The second way proposed by R.~Conte and M.~Musette, is the following:

1) Choose a positive integer $m$ and define the first order ODE
   (6), which contains unknown constants $a_{jk}$.

2) Compute coefficients of the Laurent series solutions for~$(2)$ or $(3)$
   with some fixed $C$.
   The number of coefficients  has to be greater than the number of
    unknowns.

3) Substituting the obtained coefficients, transform eq.~$(8)$ into
   a linear and overdetermined system in $a_{jk}$ with
   coefficients depending on arbitrary parameters.

4) Eliminate the $a_{jk}$ and obtain the nonlinear system in
   five parameters.

5) Solve the obtained system.

To obtain the explicit form of the elliptic function, which satisfies the
known first order ODE, one can use the classical method due to Poincar\'e,
which has been implemented in Maple~\cite{Maple} as the package
"algcurves"~\cite{MapleAlgcurves}.

The second way has a few preferences. The first preference is that one
does not need to transform system~$(2)$ to one differential equation
either in $y$ or in $x$.  Moreover at $C=-16/5$ not $x$, but $x^2$ may be
an elliptic function. To construct the Laurent series for $x^2$ is easier
than to find the fourth order equation in $x^2$.  The main preference of
the second method is that the number of unknowns in the resulting
algebraic system does not depend on number of coefficients of the first
order equation.  For example,  eq. $(6)$ with $m=8$ includes 60 unknowns
$a_{jk}$, and it is not possible to use the first way to find similar
solutions. Using the second method we obtain (independently of the value of
$m$) a nonlinear algebraic system in five variables:  $\lambda_1$, $\lambda_2$, 
$H$ and two arbitrary coefficients of the Laurent-series solutions.

The first way also has one important preference. It allows to obtain
solutions for an arbitrary $C$, whereas using the second method one has to
fix value of $C$ to construct the Laurent series solutions, because  the
resonance structure depends on $C$.

\subsection{The Computer Algebra Algorithm}

Let us consider computer algebra procedures, which assist
to construct the first order equation in the form $(1)$ with
the given Laurent-series solutions:
$$
  y=\sum_{k=-p}^\infty c(k)t^k,
$$
where $p$ is some integer number. We can eliminate from eq.~$(1)$ terms
more singular than~$y_t^m$:
$$
F(y_t,y)\equiv\sum_{k=0}^{m}
\sum_{j=0}^{j<=(m-k)(p+1)/p}a_{jk}^{\vphantom{27}}\: y^j y_{t}^k=0,
\qquad a_{0m}^{\vphantom{27}}=1. \eqno(4)
$$
At singular points $y_t^m$ tends to infinity as $t^{m(p+1)}$, so we can
present $F(y_t,y)$ as the Laurent series, beginning from this term:
$$
  F(y_t,y)=\sum_{s=-m(p+1)}^{N_{max}} K_st^s \eqno(5)
$$
and transform $(4)$ in overdetermined algebraic system: $K_s=0$ in
$a_{ij}$.  We choose $N_{max}$ to be more than the number of coefficients
$a_{ij}$.  The Maple procedures, which make this transformation are
presented in Appendix. The trivial variant is the following:  Procedure
$quvar(m,p)$ calculates the number of coefficients $a_{ij}$, procedure
$equa(a, m, p, yp2, dyp2)$ constructs the first order equation in the form
(4) and procedure $equalaur(a, m, p, Nmax)$ constructs the Laurent series
$(5)$.  $Nmax$ should be more than $quvar(m,p)$.  The first computer
algebra realization, which generates only terms used in future, has been
written in AMP~\cite{AMP} by R.~Conte.  This algorithm bases on the
$\alpha$--method of the Painlev\'e test.  Our realization bases on
transformations of the Laurent series and generates only useful terms as
well. There exist Maple~\cite{Maple} and REDUCE~\cite{REDUCE} realizations
of our algorithm.  The Maple realization \{procedure
$equlaurlist(a,m,p,ove,c)$\} is presented in Appendix.

Let us consider how this procedure works.  We put $a(0,m)=1$, other
$a(i,j)$ are unknown.  The procedure $equlaurlist(a,m,p,ove,c)$ does the
following:

1) Calculates $Nmax:=quvar(m,p)+ove;$

2) Constructs the list which corresponds to eq.~$(5)$:
   $fequlist:=equalist(a,m,p);$
   For example, if $m=2$ and $p=2$ we obtain
$$
   fequlist:=[[a[0,0], 0, 0], [a[1,0], 1, 0], [a[2,0], 2, 0], [a[3,0], 3,
0], [a[0,1], 0, 1], [a[1,1], 1, 1], [1, 0, 2]];
$$

3) To simplify the following procedures puts
$$
  \forall k=-m(p+1)..-p-1 : c(k)=0;
$$

4) Constructs the list of the Laurent series coefficients of $F(y_t,y)$
   ($laurlist$). Coefficient corresponding $t^k$ is constructed due to
   procedure $oneequlaur(c,fequlist,k,p)$ as the sum of the corresponding
   coefficients of terms $a[i,j]y^jy_t^j$. These coefficients are
   calculated by procedure \linebreak $monomlaur(c,mon,k,p)$, where
$mon=[a[k,j],i,j]$.

The obtained system is linear in $a[i,j]$ and nonlinear in parameters of
the Laurent series. This system can be transformed into a nonlinear system
in parameters of the Laurent series, so the number of unknowns does not
depend on $m$. The resulting nonlinear system can be solved using the
standard Gr\"obner basis method.

\section{CONCLUSION}

The Painlev\'e test is a very useful tool to find single-valued solution.
The corresponding computer algebra algorithm has been constructed in
Maple and REDUCE. The "naive" algorithm calculates many terms to be
discarded on the following step. Our algorithm calculates only useful
terms.

The author is grateful to \  R.~Conte \  and \  V.~F.~Edneral \
for valuable discussions.  This
work has been supported by Russian Fede\-ration President's
Grants NSh--1685.2003.2 and NSh--1450.2003.2 and by the grant of  the
scientific Program "Universities of Russia" 03.02.028.

\section*{Appendix}

\begin{verbatim}


quvar:=proc(m::integer, p::integer)

# 10.10.2003
# This procedure calculates the number of terms in first order autonomous 
# ODE, which solutions tend to infinity as 1/t^p. 
# The maximal degree of the derivative is equal to m.

local k, j, numterm;
numterm:=0;
for k from 0 to m-1
    do for j from 0 while p*j <= (p+1)*(m-k)
       do numterm:=numterm+1;
       od;
    od;
return numterm;
end;


equa:=proc(a, m::integer, p::integer, yp2, dyp2)

# 10.10.2003
# This procedure constructs the first order autonomous polynomial ODE, which
# solutions tend to infinity as 1/t^p. 
# The maximal degree of the derivative dyp2 is equal to m.

local equ, k, j, numterm;
equ:=0;
for k from 0 to m
    do for j from 0 while p*j <= (p+1)*(m-k)
       do equ := equ+a[j,k]*yp2^j*dyp2^k;
       od
    od;
return equ;
end;


equalaur:=proc(a, m::integer, p::integer, Nmax::integer)

# 10.10.2003
# This procedure expands the first order polynomial autonomous ODE in
# the Laurent series, including terms from 1/t^p to t^Nmax.
# The maximal degree of the derivative dyp is equal to m.

local max,equ,k,j,y,dy,equlist,t;
equ:=equa(a,m,p,yp,dyp);
y:=0;
for k from -p to Nmax-p do y:=y+c(k)*t^k od;
dy:=diff(y,t);
max:=quvar(m,p)+1;
if Nmax > max then max:=Nmax fi;
for k from 0 to m
  do
    dyp(k):=convert(taylor(eval(dy**k*t^((p+1)*m)),t,max),polynom)
  od;
for k from 0 to iquo(m*(p+1),p)
  do
    yp(k):=convert(taylor(eval(y**k*t^((p+1)*m)),t, max),polynom)
  od;
equlist:=[];
equ:=expand(eval(equ*t^(-(p+1)*m)));
for k from 1 to max
  do
    equlist:=[op(equlist),asubs(t=0,equ)];
    equ:=diff(equ,t)/k;
  od;
return equlist;
end;



equalist:=proc(a, m::integer, p::integer)

# 30.10.2003
# This procedure constructs the first order autonomous ODE is a list.
# solutions tend to infinity as 1/t^p. 
# The maximal degree of the derivative is equal to m.

local fequlist, k, j;
fequlist:=[];
for k from 0 to m
    do for j from 0 while p*j <= (p+1)*(m-k)
       do fequlist:=[op(fequlist),[a[j,k],j,k]];
       od
    od;
return fequlist
end;


ydegree:=proc(c,n,j,p)

# 1.11.2003
# This procedure constructs the Laurent series for y^n.
# Solutions tend to infinity as 1/t^p. mon:=[a[i,j],i,j].

local sumy,k;
if n=1 then return c(j)
   else sumy:=0;
        for k from -p to j+p*n
        do sumy:=sumy+c(k)*ydegree(c,n-1,j-k,p);
        od;
        return sumy;
fi;
end;


dydegree:=proc(c,n,j,p)

# 1.11.2003
# This procedure constructs the Laurent series for dy^n.
# Solutions tend to infinity as 1/t^p. mon:=[a[i,j],i,j].

local sumdy,k;
if n=1 then return (j+1)*c(j+1)
   else sumdy:=0;
        for k from -(p+1) to j+(p+1)*n
        do sumdy:=sumdy+(k+1)*c(k+1)*dydegree(c,n-1,j-k,p);
        od;
        return sumdy;
fi;
end;


monomlaur:=proc(c,mon,j::integer,p::integer)

# 1.11.2003
# This procedure constructs the Laurent series for the monomial 
# a[i,j]*y^i*dy*j.  
# Solutions tend to infinity as 1/t^p. mon:=[a[i,j],i,j]

local k,coef,ydeg,dydeg,sum;
coef:=op(1,mon);
ydeg:=op(2,mon);
dydeg:=op(3,mon);
if ydeg=0 then
  if dydeg=0 then
     if j=0 then return coef
     else return 0
     fi;
  else return coef*dydegree(c,dydeg,j,p)
  fi;
else if dydeg=0 then return coef*ydegree(c,ydeg,j,p)
     else sum:=0;
          for k from -p*ydeg to j+(p+1)*dydeg
             do sum:=sum+ydegree(c,ydeg,k,p)*dydegree(c,dydeg,j-k,p);
             od;
          return coef*sum;
     fi;
fi;
end;


oneequlaur:=proc(c, fequlist,  j::integer, p::integer)

# 7.11.2003
# This procedure constructs the j-th term of the Laurent series of the
# first order autonomous ODE (the list fequlist).
# solutions tend to infinity as 1/t^p.

local equj,k,test;
equj:=0;
test:=nops(fequlist);
for k from 1 to nops(fequlist)
   do equj:=equj+monomlaur(c,op(k,fequlist),j,p); od;
return equj;
end;


equlaurlist:=proc(a,m::integer,p::integer,ove::integer,c)

# 10.11.2003
# This procedure constructs the Laurent series of the
# first order autonomous ODE with maximal degree of the derivative is equal to m.
# Solutions tend to infinity as 1/t^p.
# c(k) are the Laurent series coefficients of the function.
# The length of the resulting list is quvar(m,p)+ove.

local k, laurlist, fequlist, Nmax;
Nmax:=quvar(m,p)+ove;
fequlist:=equalist(a,m,p);
for k from -m*(p+1) to-p-1 do c(k):=0 od;
laurlist:=[];
for k from -m*(p+1) to Nmax-m*(p+1) do
     laurlist:=[op(laurlist),oneequlaur(c,fequlist,k,p)] od;
 return laurlist;
end;

\end{verbatim}
\end{document}